\pdfoutput=1
\documentclass[sigconf]{acmart}
\usepackage{booktabs}
\usepackage[utf8]{inputenc}
\usepackage{appendix}
\usepackage{graphicx}
\usepackage{balance} 
\usepackage{color}
\usepackage[utf8]{inputenc}
\usepackage{listings} 
\usepackage{comment}
\usepackage{tabularx} 
\usepackage{paralist}
\usepackage{subfig}
\usepackage{hyperref}
\usepackage{amsmath}
\usepackage{amssymb}

\hypersetup{
  pdfkeywords={}
}

\acmConference[CASTrm]{Conference of Aspiring Students in Tech Rhein-Main}{March 2019}{Darmstadt, Germany}
\acmYear{2019}
\copyrightyear{2019}

\begin{document}

\title{\texorpdfstring{Impersonating LoRaWAN gateways using Semtech Packet Forwarder}{A hypothetical proposal on realising acknowledgement spoofing in LoRaWAN}}

\author{\texorpdfstring{Lukas Simon Laufenberg}{Lukas Simon Laufenberg}}

\affiliation{
  \institution{Secure Mobile Networking Lab, Technische Universtit\"at Darmstadt}
  }
\email{llaufenberg@seemoo.tu-darmstadt.de}

\begin{abstract}
Low Power Wide Area Network (LPWAN) technologies like the Long Range Wide Area Network (LoRaWAN) standard provide the foundation of applications realizing communication and intelligent interaction between almost any kind of object. These applications are commonly called Smart Cities and the Internet of Things (IoT). Offering the potential of great benefits for mankind, these applications can also present a significant risk, especially when their security is compromised. This paper's work analyzes the possibility of two particular scenarios of impersonating a LoRaWAN gateway combining existing attacks. Impersonated gateways are of use when exploiting vulnerabilities already shown by other researchers. We give a basic overview about LoRaWAN, the Semtech Packet Forwarder protocol, attacks needed to perform the impersonation, and assumptions made. We explain our attack and propose countermeasures to increase the security of LoRaWAN networks. We show a gateway impersonation is possible in particular circumstances but can be detected and prevented.
\end{abstract}

\maketitle

\section{Introduction}

Services and applications making use of data from and the interaction of everyday objects and devices, not necessarily electronic ones, grow in importance. The research and advisory company Gartner Inc. predicts that the number of objects or devices connected to the Internet will reach 25 billion by 2020\cite{GartnerIOT}. Smart City and IoT are terms, which describe the goals behind using these applications and services. LPWAN technologies like LoRaWAN provide the fabric to solutions implementing smart grids, sensor networks, and other applications realizing these goals. SK Telecom, a telecommunication company in South-Korea offers a nationwide LoRaWAN network\cite{pulsenewsSKTelecom}. The crowd-sourced project called The Things Network (TTN) offers a public, international, and non-commercial LoRaWAN. Hence we can gather LoRaWAN will affect the everyday life of many people in the future and its security therefore is of great importance for society.\\
There has been security centered research on many parts of LoRaWAN, e.g. attacking communication and availability of end devices. One attack we find in this research is ACK spoofing as proposed by Yang\cite{YangLoRaSec}. This attack proves that gateways can be used to pretend the reception of messages without actually forwarding them. It uses a malicious gateway, controlled by an attacker.
However, only briefly indications about adding a malicious gateway or infiltrating present gateways are presented to achieve them. The aim of this analysis is to propose and examine a new approach enabling an attacker to use another gateway to mimic the behavior of a non malicious one, while disabling the real gateway. This approach consists of two scenarios, where a gateway is disabled, either by disconnecting it, or by leveraging the jamming with commodity LoRaWAN hardware as shown by Aras et al\cite{SelJam}. We then use limitations, traces of the attacks, and common security technologies to propose countermeasures against the attack.\\
As LoRaWAN offers flexibility in the use of upper layer protocols, a scenario has to be specified resembling existing applications. For gateway to server communication the Semtech Packet Forwarder Protocol is used and an example gateway registration process is derived from TTN, explained on the organization's website.\\
To investigate the possibility of a gateway impersonation we combine the works of Yang\cite{YangLoRaSec}, Miller et al.\cite{MillerSecureLoraWP}, and Aras et al.\cite{SelJam} and extend them with examining the specifications of LoRaWAN \cite{LoRaWANSpec} and the Semtech Packet Forwarder Protocol \cite{SemtechUDP} as well as instructions publicly available on TTN's websites. The results though will have to be tested in practical conditions to evaluate effort and probability of success.\\
This paper consists of 5 parts. We give a brief overview about the LoRaWAN protocol and the LoRa modulation in Section 2. Since the LoRaWAN specification does not define how gateways should be connected to network servers, we describe the Semtech Packet Forwarder Protocol and an example gateway registration process as well. In Section 3 we summarize different attacks done by other researchers. First we take a look at attacks making use of malicious gateways. Then we describe how messages can be jammed, recorded and resent later or at other locations. In Section 4 we describe our impersonation approach. Further we propose two possible ways to disable legitimate gateways and explain their impersonation step by step. Resorting to findings in Sections 4 we recommend countermeasures against gateway impersonation in Section 5. Concluding, we summarize this papers findings and approaches for further investigations in Section 6.

\section{Background}

This section gives an introduction to LoRaWAN and related technologies, needed to understand the attack approach.
These include the LoRa modulation scheme, the gateway registration process from TTN and the Semtech Packet Forwarder Protocol.

\subsection{LoRaWAN\textsuperscript{TM}}

LoRaWAN is an open specification developed by the LoRa Alliance. It defines a communication protocol for a scenario involving devices distributed over a large area with the constraint of minimizing the end device's energy consumption, enabling devices to last on regular batteries for years. The constraint results in smaller data rates between 0.3 and 50 Kilobit per second, useful in situations where mostly sensory and controlling communication takes place. This is the case for many applications like Smart Cities and industrial automation. While the protocol is open to public development, LoRa, the physical modulation used for transmitting LoRaWAN radio messages, is intellectual property of Semtech.\\
\begin{figure}[h]
	\includegraphics[width=\linewidth]{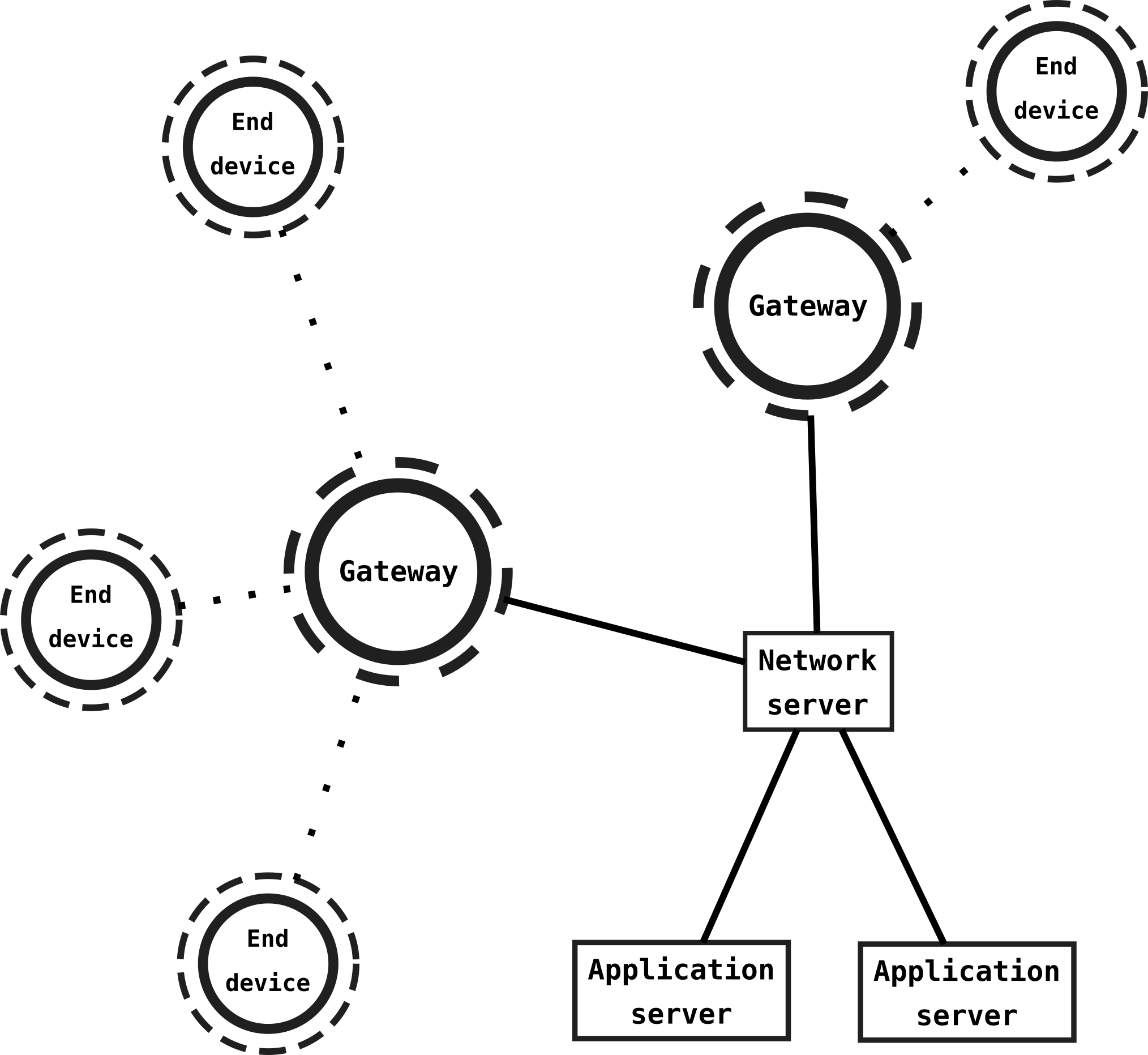}
	\caption{LoRaWAN network topology}
\end{figure}
LoRaWAN's network topology follows a star of stars pattern, as shown in Figure 1. The network server, as a central point, is the hub, all LoRaWAN data is exchanged over. On the one side it is connected to servers running applications using data from a LoRaWAN. These do not have to be physical servers but could also be virtual machines or applications running on the network server. On the other side it communicates to remote base-stations called gateways. How these connections are accomplished is not part of the specification, suggesting commonly used IP connections and related upper layer protocols for this purpose. Gateways are stationary LoRa-transceivers receiving uplink radio messages from the end devices and sending the demodulated data to the network server. These transmissions are mostly data for applications and requests to join a network. Gateways are also used for downlink messages from application or network servers, needed for configuring devices. Although LoRaWAN communication is bidirectional, uplink traffic from end devices to application servers is considered predominant.\\
The specification offers different modes of operation, which can present another task for gateways. To reduce power consumption, devices in Class A networks have only two short receiving windows after transmitting an uplink message, offering the gateway an opportunity to transmit downlink messages to a device. In Class B networks a gateway regularly sends out time synchronized beacons, which open additional device receiving windows. In Class C networks, continuous reception of messages by devices is possible, increasing the power consumption and leading to a corresponding decrease of lifespan.\\
To establish confidentiality of the data and integrity of the messages, cryptography is used. At first an application key is created on the application server and placed on every device destined for the same. Two session keys are derived from it, when a device enters the network. One serves as the application session key protecting the confidentiality of the message payload between end device and application server. The other is used as network session key protecting the integrity of messages between end device and network server, respectively present on both.\\
Nonces, the concept of numbers only used once, are utilized in various cases, e.g. sequential counters for uplink and downlink messages, preventing replay attacks in certain scenarios, but also offering possibilities for denial of service attacks during the network joining procedure, as shown Tomasin et al\cite{JoinLoRa}.\\
The radio channels used for LoRaWAN communication are located in different ISM-Bands depending on the regulation authority responsible for the network's area. Resulting from this, parameters like transmission power and duty cycle differ. For this paper we assume European regulations, though the findings apply for other regulation areas as well. Aside from this, network provider policies and the LoRaWAN specification define additional limits for certain aspects of the protocol\cite{TTNdutycycle}. Since the ISM-Bands require no licensing fees for using, there can be interference caused by other radio signals. Therefore LoRaWAN makes use of cyclic redundancy checks (CRC) to protect the physical integrity of the payload\cite{LoRaWANSpec}.

\subsection{The LoRa modulation}
The modulation used for LoRaWAN is based on Chirp spread spectrum (CSS), i.e. the modulation uses the whole bandwith of the communication channel by encoding a message in sequences of chirps, which are either ascending or descending linear changes in frequency. This makes it resistant to noise, which is important when communicating in ISM bands, as these are less regulated, leading to more communication and possible interference\cite{LoRaDG}.\\
Also colliding messages do not necessarily lead to message loss, but can be received simultaneously with different spreading factors\cite{SelJam}.\\
The spreading factor describes how many physical elements are used for transmission of a symbol. This means, a higher spreading factor uses more physical space, e.g. bandwidth or as in LoRa's case time duration, to transmit the same information. The spreading is what makes a message more resistant to interferences, but also reduces the transmission speed as a message will be longer.\\
To get a grasp of how the chirps in LoRa messages are structured, a visual representation of the frequency spectrum can be seen in Figure 2 created by Knight and Seeber\cite{declora}.\\
\begin{figure}[h]
	\includegraphics[height=\linewidth]{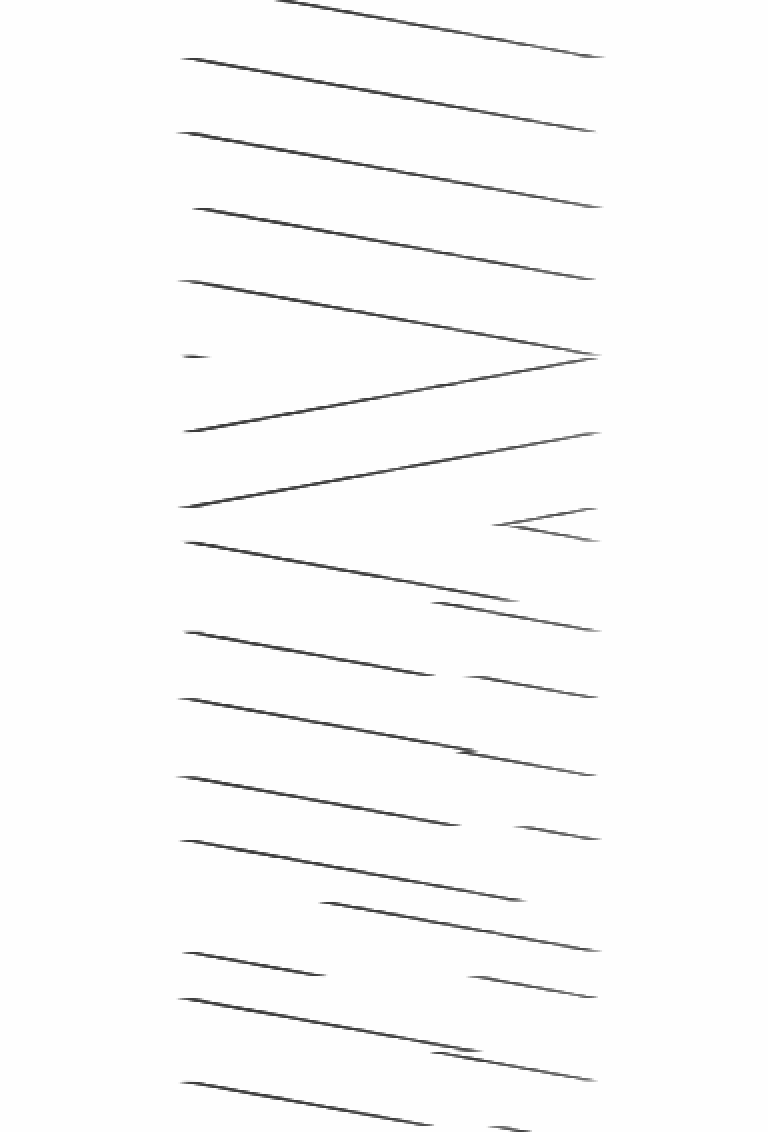}
	\caption{Spectrogram excerpt of a LoRa chirp y-axis: Time, x-axis: Frequency | taken from \cite{declora}}
\end{figure}

\subsection{Gateway registration}

Registering gateways to networks is not a necessity demanded by LoRaWAN. For example TTN is accepting traffic even of unregistered Semtech Packet Forwarder Gateways. However in this case the gateway and traffic are marked as untrusted. Registering gateways is a common practice, since it can provide an overview of the network, but also be necessary if gateways use more complex packet forwarders requiring configuration, e.g. depositing cryptographic keys. For this paper a registration process that is similar to TTN's is presumed. After creating an account on the organization's website, a gateway can be registered by entering data into a form. The requested information includes the geographic location, whether the Antenna is placed indoors or outdoors, the frequencies used, a short description, the unique gateway identifier and the protocol that is used for packet forwarding, i.e. the transmission of radio packets from gateway to network server\cite{TTNreg}.\\
The organization supports two types of protocols. Packet Forwarders using the Gateway Connector Protocol, developed in parallel with the TTN Packet Forwarder, have to be registered since they require further configuration. Although the protocol is still supported, by the time of writing this paper the development of the packet forwarder was put on hold \cite{TTNdevstop}.\\
The other protocol supported is the Semtech Packet Forwarder Protocol. When registering a gateway of this type, a Device EUI of 8 bytes length can be taken from the gateways LoRa module, or generated otherwise. It then has to be entered during registration on the website as well as the local configuration file of the Semtech Packet Forwarder or another using the same protocol. After configuring location information, the gateway is visible on the official map on the website. When registration is finished a gateway can send traffic that is not marked untrusted, using the registered EUI\cite{TTNreg}.

\subsection{Semtech Packet Forwarder Protocol}

Packet forwarders are applications running on gateways. They take care of sending demodulated radio packets arriving at the gateway to a network server and transmitting packets coming from the same via a built in LoRa transceiver. The LoRaWAN specification does not provide a protocol or specific implementation to use, since the scope of LoRaWAN is to provide a flexible foundation to applications.\\
The protocol for Semtech Packet Forwarder is historically the first gateway connection protocol and was developed by Semtech Corporation who maintain it. The Semtech Packet Forwarder Protocol is not recommended for production use and explicitly developed without authentication \cite{SemtechUDP}. However this packet forwarder is still used in many commodity gateways \cite{TTNsemtechUDP}. TTN offers a public map of all gateways deployed in their networks, showing the devices names \cite{TTNhp}. Given the practice that names of devices using this protocol are visible on the map with prefix "eui-" and the gateway identifier, many gateways using this protocol can be seen, despite being considered unsafe by Semtech Corporation and TTN\cite{TTNreg}. Current gateway data from TTN used for generating the map can be obtained in JSON format from the url "https://www.thethingsnetwork.org/gateway-data/". When accessing the data on 23rd March 2019, of 6869 gateway entries present in the dataset, which had been last seen after 20th March 2019, 4968 had an "eui-" prefix. Given this information, it can be assumed the majority of gateways in TTN are using Semtech's protocol. We can therefore reason, an attack requiring the assumptions made is relevant in real world scenarios. At the same time, the website stated there were "6646 gateways up and running", so there might be additional factors TTN uses for detecting the functionality of a gateway.\\
Semtech's packet forwarding is based on the connectionless User Datagram Protocol (UDP). Messages are sent to the application listening on the server with IP-address and port number specified in the destination fields of the datagram. Also source fields with IP-address and port number of the sending application exist. There are no mechanisms in the form of acknowledgment or retransmission to avoid packet loss. Therefore applications using the protocol should be anticipating data loss and prepare for it \cite{UDPspecs}.\\
Since LoRaWAN traffic is suspected to be mostly consisting of uplink traffic, it comes as no surprise that uplink and downlink messages are handled differently in the Semtech Protocol\cite{LoRaWANSpec}.\\
\subsubsection{Uplink communication}
\begin{figure}[h]
	\includegraphics[width=\linewidth]{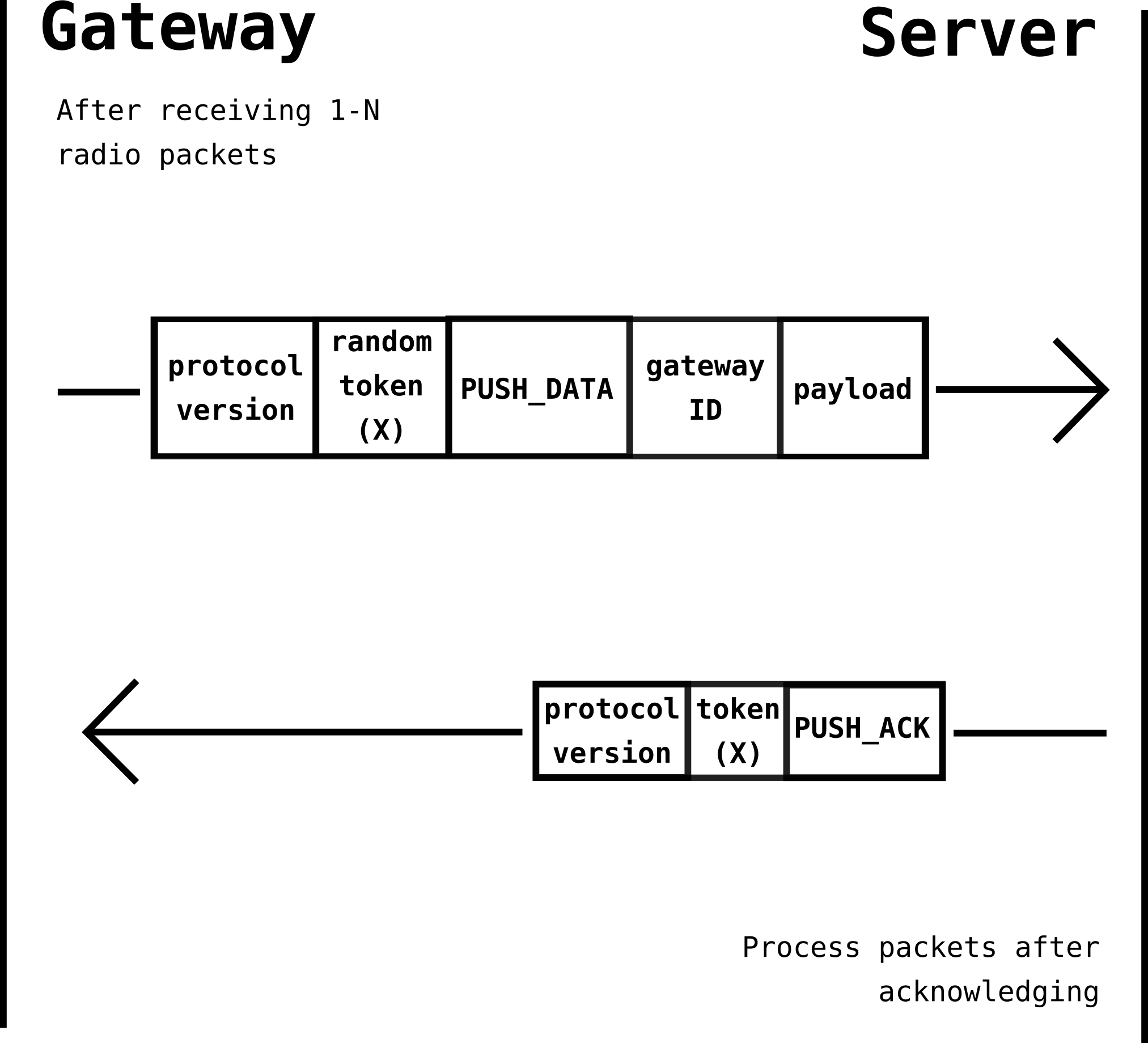}
	\caption{Packet Forwarder Protocol uplink sequence diagram}
\end{figure}
For an uplink message to be sent, a radio packet addressed to the gateway must arrive at the gateway's transceiver. After a specified number of packets arrived, the forwarder initiates communication by sending a PUSH\_DATA message to the server, composed as seen in Table 1.
\begin{table}[h]
	\begin{tabular}{r | l} 
		Bytes & Function \\
		\hline
		0 & protocol version \\
		1-2 & random PUSH\_ACK token \\
		3 & PUSH\_DATA identifier (0x00) \\
		4-11 & Gateway identifier \\
		12-end & JSON-like payload \\
	\end{tabular}
	\caption{PUSH\_DATA message}
\end{table}
The gateway's identifier is supposed to be the 64 Bit EUI from the LoRa transceiver module on the gateway but can be set to any 8-byte string in the gateway's configuration. The payload is encoded in base64 and encapsulated in a JSON-like object containing status information as well. Of special interest is the "stat" field, presenting information about the CRC checksum. This field can have the values 1, for a correct checksum, -1, for an incorrect one, and 0 if no checksum is used.\\
If the PUSH\_DATA message is received by the network server, a PUSH\_ACK message is returned containing the PUSH\_ACK token. This is only used for network quality measurements, packets are not retransmitted if no PUSH\_ACK message arrives at the gateway \cite{SemtechUDP}. This process is illustrated in Figure 3.\\

\subsubsection{Downlink communication}

As gateways can be connected to network servers with a lot of different technologies, it is probable that some connections involve network address translation, which keeps a server from initially sending a message to the gateway. However since this behavior is needed, the protocol regularly sends PULL\_DATA messages to the server to open a route to the gateway and at the same time informing the server about the open route. The structure of these messages can be seen in Table 2.
\begin{table}[h]
	\begin{tabular}{ r | l} 
		Bytes & Function \\
		\hline
		0 & protocol version \\
		1-2 & random token \\
		3 & PULL\_DATA identifier (0x02) \\
		4-11 & Gateway identifier \\
	\end{tabular}
	\caption{PULL\_DATA message}
\end{table}
After acknowledging the PULL\_DATA message the server is now able to send PULL\_RESP packets downlink. Theses packets include the physical payload that the gateway transmits to the end device via transceiver. If a transmission to the gateway is successful, the gateway responds with a TX\_ACK message, shown in Table 3, optionally containing a JSON object with status information about the transmission. 
\begin{table}[h]
	\begin{tabular}{ r | l} 
		Bytes & Function \\
		\hline
		0 & protocol version \\
		1-2 & same token as the PULL\_RESP packet \\
		3 & TX\_ACK identifier (0x05) \\
		4-11 & Gateway identifier \\
		12-end & JSON status-payload \\
	\end{tabular}
	\caption{TX\_ACK message}
\end{table}

A sequence diagram showing the explained excerpt of downlink communication is shown in Figure 4.

\begin{figure}[h]
	\includegraphics[width=\linewidth]{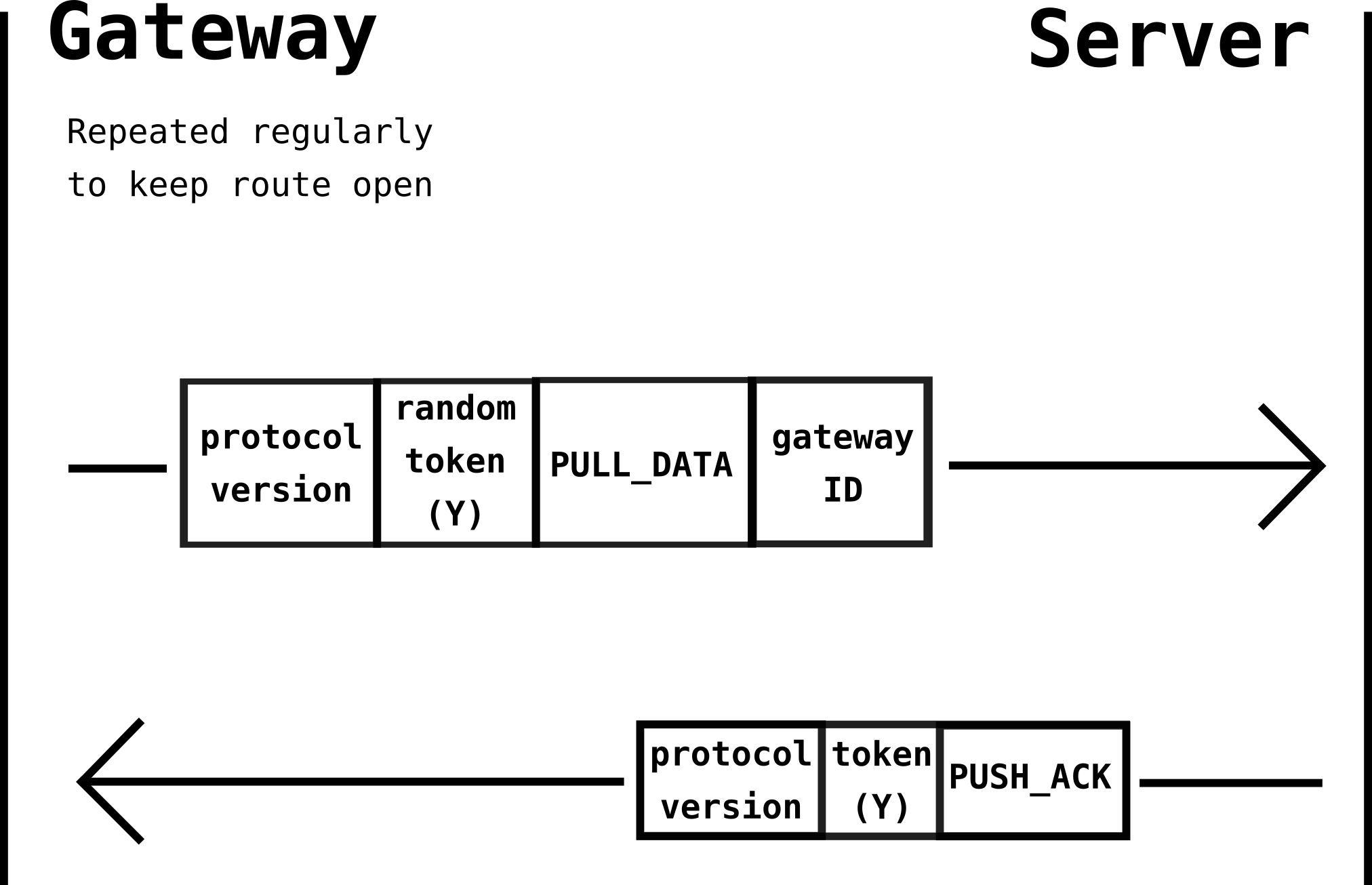}
	\includegraphics[width=\linewidth]{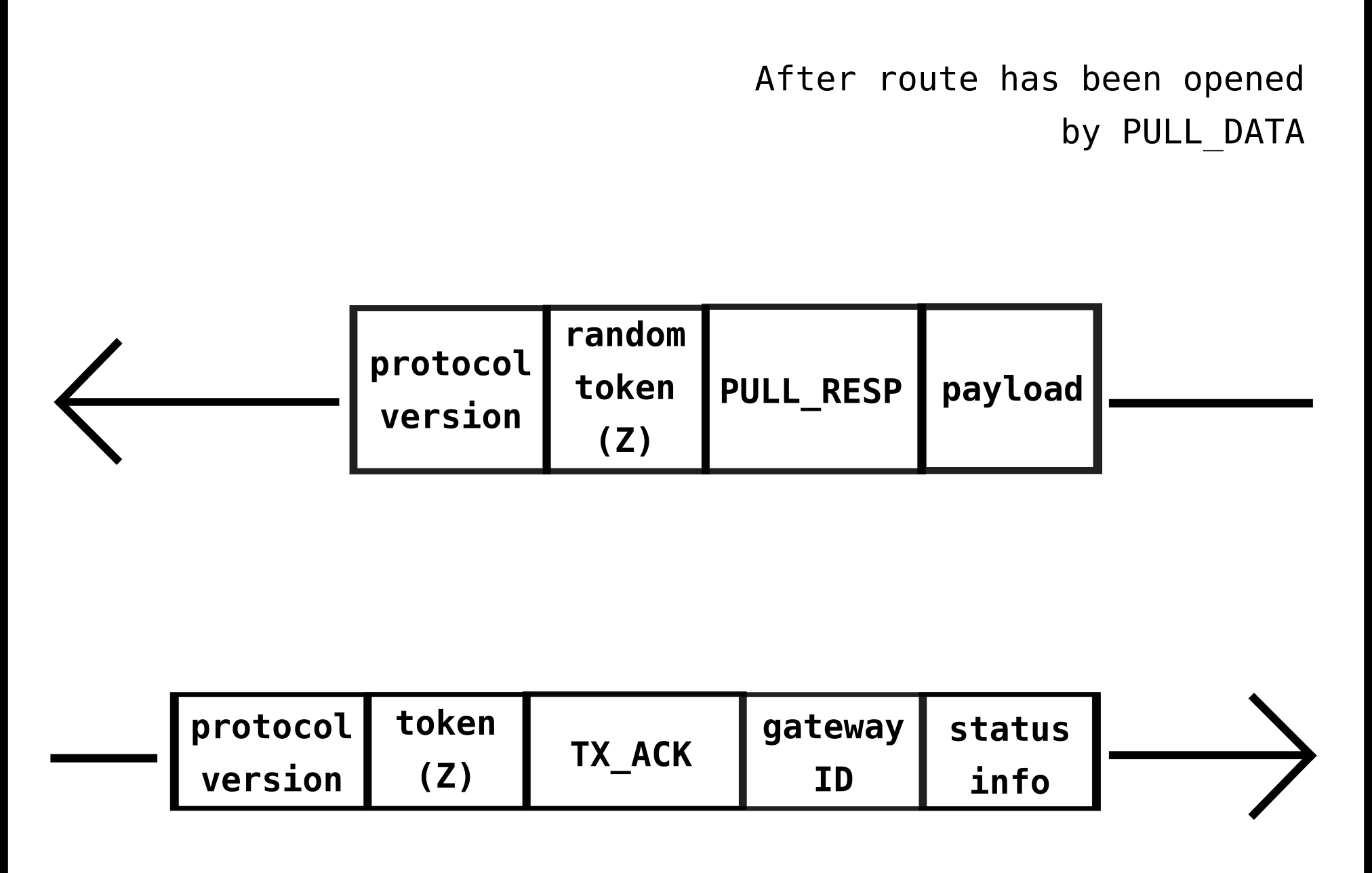}
	\caption{Packet Forwarder Protocol downlink sequence diagram}
\end{figure}

\section{Related Work}

In this section we describe two attacks which a gateway under an attackers control can realize. Afterwards we will provide a short summary of the attacks we leverage to disable gateways before impersonating.

\subsection{Malicious Gateways}

An attacker having control over gateways poses a threat to LoRaWANs, as every gateway includes a transceiver enabling it to constantly listen for messages and sending the same as well. These capabilities already enable a lot of attacks, where these technical capabilities are required. Some of these attacks are described in the works of Yang \cite{YangLoRaSec}, and Aras et al\cite{SelJam}. However there are two attacks, where the role of a gateway is required for the attack, or at least suggests itself.

\subsubsection{Beacon spoofing}

Robert Miller describes an attack that can be achieved just by a  using a regular transmitter capable of sending LoRa Messages\cite{MillerSecureLoraWP}. For the attack to work it is however necessary a Class B LoRaWAN is used, which provides a new task to gateways. We explain millers attack therefore with an attacking gateway, although regular end devices could be used as well, since the beacons are neither integrity protected, nor signed and can therefore be spoofed by any attacker able to send high power LoRa messages.\\
When a network is running in Class B mode of operation, the gateway sends out beacons, small radio packets presenting time synchronization and location information. It lets end devices open additional receiving windows for downlink messages in regular intervals, despite the two windows that are already opened after an uplink message. This decreases delays, as downlink messages can be sent more frequently.\\
By sending out beacons randomly a malicious gateway could desynchronize an end device from receiving windows of another gateway. This could cause a denial of service, as the legitimate gateway sends messages when the end device is not receiving. This can also be adapted to increase power consumption of end devices and lowering their lifespan. In this case, the beacon frequency is increased, opening more receiving windows than necessary, wasting energy. Even indirect manipulation of data is possible with the transmission of wrong location data in the false beacon especially relevant, if the end device is not stationary\cite{MillerSecureLoraWP}.

\subsubsection{ACK spoofing}

The arrival of LoRaWAN uplink messages at the network server is not confirmed to the end device by default. In certain applications it is useful to have these kind of acknowledgments for end devices, so packet loss can be detected and messages can be retransmitted. To make this possible, LoRaWAN has so called uplink confirmed messages, that trigger the network server to send a packet to the end device acknowledging the reception of the message. This packet is called ACK. In Table 4, the messages structure of a LoRaWAN message is shown. In the frame control byte the third most significant bit is set, if the message is an acknowledgment\cite{LoRaWANSpec}.\\
\begin{table}[h]
	\begin{tabular}{ r | l} 
		Bytes & Function \\
		\hline
		0 & message header \\
		1-4 & device address \\
		5 & frame control byte \\
		6-7 & frame counter \\
		8-11 & message authentication code \\
	\end{tabular}
	\caption{LoRaWAN ACK message}
\end{table}
For the following attack, the frame counter (FCnt) is of interest. Its purpose is to indicate, that the last message sent has arrived at the server and it is increased every time, an ACK is sent down to the specific device. The ACK is only accepted, if FCnt is larger than the last one, the device received. As FCnt acts as a sequential Nonce and a message authentication code, created with the network session key, is attached, an attacker can neither create a valid new ACK, nor replay an already used one.\\
Yang proves this to be a design flaw, posing a security threat if an attacker has control over a gateway\cite{YangLoRaSec}. This attack abuses the fact, that FCnt does not specify the message that should be acknowledged. In case of an attacker controlling the gateway, the ACK for a message can be held back at the gateway. Since the end device now believes the message got lost, it will retransmit the message a few times, but finally assume the message got lost. The next message, the device is requesting an acknowledgment for, is now be dropped at the gateway. The gateway can now send the ACK recorded for the message before, to pretend the message was received. From the perspective of the device, the lost message is received, while the received message is lost\cite{YangLoRaSec}.\\
As of LoRaWAN version 1.1, the frame counter of the message to be acknowledged, is incorporated into calculation of the acknowledgment's message integrity code\cite{LoRaWANSpec}. This way, the spoofing is detected by failure of integrity verification. This effectively prevents the attack for all end devices operating only with this version.

\subsection{Triggered jamming, selective jamming, and wormhole attack}
We now have stated what an attacker is able to do, when in control of a gateway. In order to achieve the impersonation, additional attacks are needed. Aras et al. explain different approaches to jamming LoRaWANs using commodity LoRaWAN devices\cite{SelJam}. We use some of these approaches to develop the impersonation.
\subsubsection{constant jamming of a channel}
The easiest approach to jamming is simply placing a LoRa transmitter sending a constant signal of high power next to the target. This will constantly prevent any connection on the channel by drowning any incoming message. This however can be easily detected by network operators and prevented by switching to a different channel\cite{SelJam}.
\subsubsection{triggered jamming of a channel}
Another approach uses a utility many devices provide, although this behavior is not required by the LoRaWAN standard: To avoid collisions LoRa modules can detect, whether a message is transmitted on a channel at a point in time. The device can now use this information to send messages, when the communication channel is not occupied by the message of another device. Obviously this can be used conversely to collide a random LoRa message with a legitimate message from another device. This is harder to detect than constant jamming\cite{SelJam}.
\subsubsection{selective jamming of a device}
The idea of triggered jamming can be improved by using two LoRa devices and sophisticated scanning. One device is close to the sender, scanning the message's Frame Header for the Device Address (DevAddr) of the targets. When a message with a targets DevAddr is received at the scanning device, a faster communication channel is used to trigger the second device close to the receiver, to start jamming the transmission by sending a high power LoRa message arriving at the receiver roughly at the same time of the actual message. Depending on the speed of the connection used between the attacking devices, this will mostly affect bits at the end of the legitimate LoRa message, where mostly checksums and message integrity codes are located. As the jamming operation only takes place when a message is sent by specific devices, it is even harder to detect than triggered jamming and can be misinterpreted as a checksum failure caused by noise or bad conditions\cite{SelJam}.
\subsubsection{selective jamming with wormhole attack}
Combining selective jamming with a wormhole attack can enable an attacker to pretend that a device is still operating correctly while it is not.
First a selective jamming attack is performed regularly only this time, the message is not only scanned but recorded. Only after successfully finishing recording the message, the scanning device triggers the jamming device over the faster connection. This sets further timing constraints as now, not only DevAddr has to be received completely, but also the rest of the message, before jamming can begin. However the attack is still feasible if bigger spreading factors are in use since longer messages resulting from a bigger spreading factor are transmitted slower. After recording the messages they can be replayed, as the validity of messages is not depending on the time they were sent. Therefore the messages could also be replayed by another device or at another gateway\cite{SelJam}.
\section{Impersonating gateways}
The goal of impersonated gateways is to pave the way for ACK spoofing described by Yang et al\cite{YangLoRaSec}. Therefore what we are trying to achieve from a security objectives perspective is, to attack the availability of the end devices. To achieve this we also attack the integrity of messages to gateways as well as the availability of gateways themselves.
The basic concept of the attack is:\\
\begin{enumerate}
	\item The attacked gateway's unique identifier is obtained.
	\item The attacked gateway is disabled, i.e. no communication from end devices via the gateway is successful anymore.
	\item The malicious gateway is configured by the attacker to use the legitimate gateway's unique identifier for communicating with the network server.
	\item The ACK spoofing attack is conducted.
\end{enumerate}

There exist several possible ways to enact each of the four steps and they are influencing each other. As a result, an attacker has to achieve several tasks in different ways, further explained in the following sections. We examine these tasks in two groups. First we take a look at possibilities to disable gateways, their prerequisites and their applicability. The choice, which way is used affects the second group of tasks. These tasks are dealing with the implementation of gateway impersonation itself and therefore they are approached separately.
\subsection{Disabling gateways}

Disabling the legitimate gateway is an important condition for impersonation for ACK spoofing attacks. If working gateways are covering the area of the target end device, the malicious one can be circumvented. Two ways are proposed to achieve the disabling of a gateway.

\subsubsection{Physical access}

A simple approach to disabling a gateway is the disconnection from its power source or network connection. This requires both knowledge about the location and access to the power source or respectively the network connection.\\
While the latter is depending on physical protection measures, which are out of this paper's scope, we can at least provide several options in our scenario to obtain location information. Besides online maps of service providers showing gateway locations as TTN is providing publicly on their website, there is also an aspect of the LoRaWAN protocol, leaking information about the location of gateways\cite{TTNhp}. Networks running in Class B mode are sending out beacons, as explained before. These contain GPS information of the gateway, in order to enable end devices to correctly adapt when switching between coverage areas of different gateways. Since beacons are not confidentiality protected, any receiver listening to the beacon frequency and able to decode LoRa messages can obtain this location information\cite{MillerSecureLoraWP}.

\subsubsection{Jamming}

Another approach of disabling a gateway is an adaption of the selective jamming attack as described by Aras et al\cite{SelJam}. Fulfilling the aim to disable the gateway, what has to be achieved in this case is the stop of uplink communication to the gateway. Later a replay of the communication at our malicious gateway has to be enacted, so it appears to the server as if the network is operating well, while actually the attacker is performing ACK spoofing or other attacks.\\
For this attack, it is not possible to use constant jamming, as this would clog the communication channel and therefore make it impossible to use the channel for recording messages.\\
When evaluating different jamming approaches triggered jamming at first comes to mind, as it is the simplest of advanced jamming techniques. Triggered jamming enables the attacker to stop all uplink communication. The concerns of detecting the triggered jamming are less grave in this scenario, despite the fact that all devices in range are jammed, since the gateway is impersonated and the jammed messages are replayed later, simulating normal behavior of the network\cite{SelJam}. Still triggered jamming probably is not the best fit in this case, as the collision of messages is caused as soon as possible and therefore the collision of the jammer might not only affect the legitimate gateway but also recordings of the attacker.\\
Hypothetically, if recording and jamming device are located far enough from each other it might be possible to use the triggered approach for the jammer, abandoning the connection between both devices. Since triggered jamming is usually more successful in disturbing data  this change in procedure could enhance the effectiveness of the attack, if at the same time the recording device is close enough to the sender\cite{SelJam}. This however would have to be tested in a realistic environment since it is highly dependent on timing and physical parameters.\\
We find that selective jamming is the best fit, as it triggers jamming only after the message was recorded successfully.\\
In case of the impersonation, the recording device can be replaced with the malicious gateway. For this purpose modification is needed to trigger the jammer after successful recording.

\subsection{Attack procedure}

The process of impersonation is depending on how the legitimate gateway has been disabled. In both situations, the essential part is to spoof the gateway's unique identifier used by the legitimate gateway which in most cases should be registered at the network provider.

\subsubsection{Obtaining gateway identifiers}

Gateway identifiers are not treated as an asset to be securely stored, as they are even presented as the device name on TTN's homepage\cite{TTNhp}. Chances are that configuration files containing legitimate gateway identifiers might also be found for example in code repositories on public networks. Public sources are the first and easiest way to obtain identifiers.\\
Otherwise, there is still the option of sniffing a message containing the identifier when it is sent to the network server by the gateway. This poses a challenge, as an attacker has to be able to eavesdrop on communication between gateway and server. This could be realized since LoRaWAN does not specify this connection. The security of the identifier during transmission is therefore directly depending on the eavesdropping countermeasures used on lower levels of the levels. E.g. if the network connection is an unsecured wireless LAN connection, the identifier can be simply recorded using software defined radio and an analysis tool for network traces. As only a single message has to be recorded, this can be achieved very easy if security measures are insufficient.

\subsubsection{Disconnected gateways}

\begin{figure}[h]
	\includegraphics[width=\linewidth]{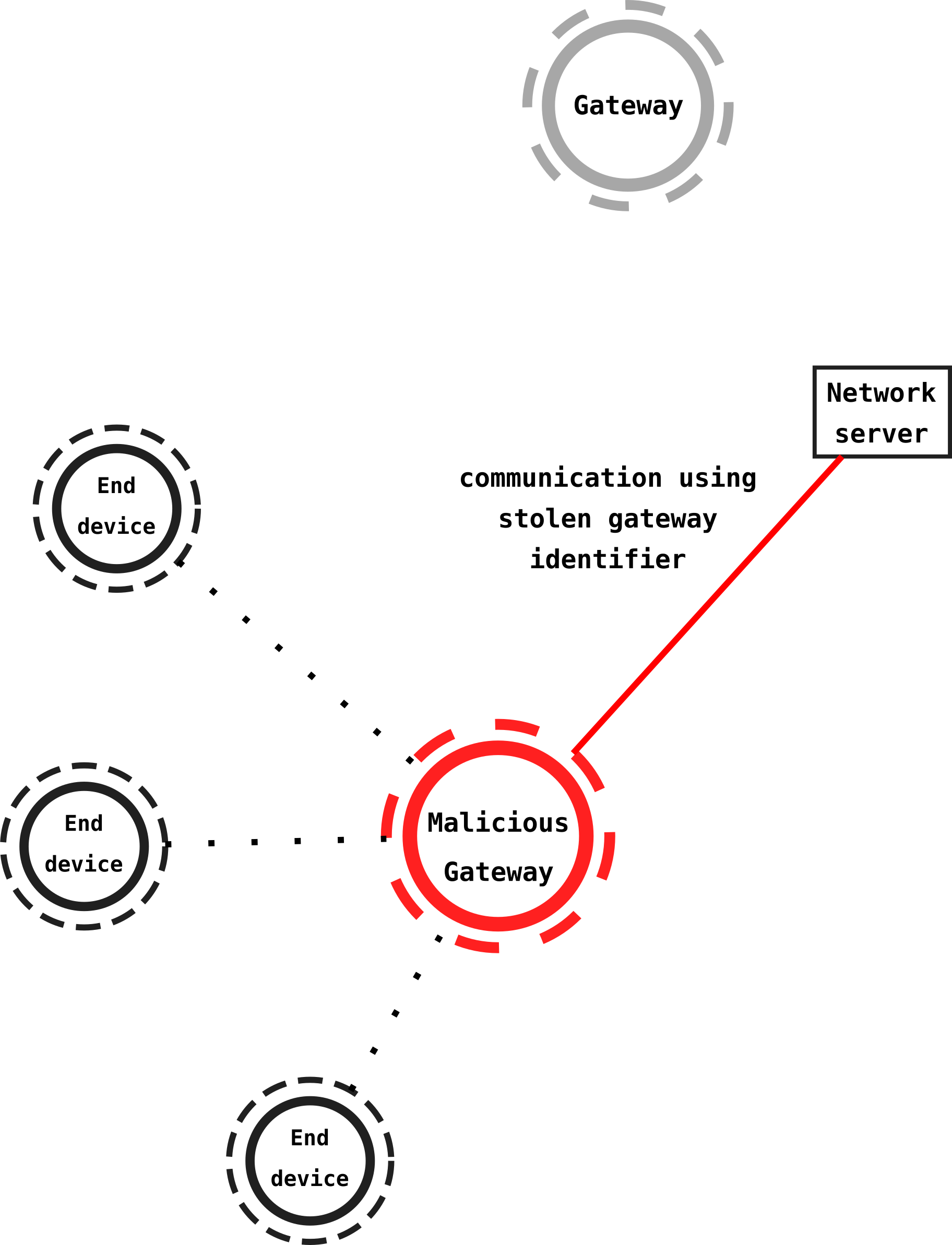}
	\caption{impersonation attack disconnecting the gateway}
\end{figure}

For our disconnection scenario, the only thing needed is a regular LoRa gateway. After obtaining the identifier, it can be configured in the gateway's local\_conf.json file. From there the packet forwarder will use the this identifier to send uplink packets and will start receiving data for the disabled gateway. After successfully disabling the legitimate gateway, there is now only one gateway transmitting the messages, appearing to the network server as legitimate. Enacting an ACK spoofing attack as explained by Yang is now possible without any limitations\cite{YangLoRaSec}.

\subsubsection{Jammed gateways}

When the communication to a gateway is jammed, it is still receptive to messages, so a simple replaying for our gateway while pausing the jamming is not possible. Choosing a similar procedure to the selective approach, where a recording device and a jammer are needed, the malicious gateway can also be used as the recording device. This comes in handy, as replaying messages can be omitted. Through this the gateway can directly serve it's purpose originally intended, sending packets to the network server. For configuration, the gateway identifier again has to be configured. However there has to be additional software installed on the gateway sending a triggering message to the jamming device. This can be omitted if it can be confirmed in practice that triggered jamming with the recording device located far enough would suffice for the attack as explained in Section 4.1.2.\\
What happens next is depending on the handling of checksum failures by the legitimate gateway. If the CRC check fails, gateways could decide to simply drop the package, reducing the number of invalid packets sent to the server. In practice it is more likely, network operators are interested in a high number of failing CRC checksums. As the Semtech Packet Forwarder Protocol can indicate a checksum failure by the 'stat' field in the payload meta data, mentioned in Section 2.4.1, in this case it appears to the server as if about half of the packets are corrupt. The intact half of the packets are sent over the attackers gateway.\\
It is of course tempting to change the UDP source address field to the original one, or apply IP spoofing, making the traffic seem to come from the legitimate gateway. Since attacks using malicious gateways, e.g. the ACK spoofing attack, are relying on receiving messages from the server, this would make it very hard to enact these attacks, as for this scenario to work, traffic to the original IP would have to be rerouted to the attacker. This would make the whole attack concept more sophisticated and less possible to realize.\\
How network servers are handling acknowledge messages sent downlink to the gateway, has to be answered by practical experiments with existing network server implementations. In this situation two gateways would be sending PULL\_DATA messages to keep a path open for the server. They appear as one from a packet forwarder protocol perspective, since the only way of identifying a gateway in the protocol is the unique identifier. The network server will have to make the choice which logical address is chosen as the destination, to send the PULL\_RESP packet containing the ACK. If the legitimate address is chosen for the downlink, the attack fails. One likely scenario is a network server choosing the IP address by using the last PULL\_DATA message that arrived at the network server. Therefore the attacker can increase the frequency of the PULL\_DATA messages. If a the server changes the downlink address after seeing a new one in the packet, chances are higher that the attackers address is chosen for sending acknowledge messages, as it appears more often than the legitimate one.

\begin{figure}[h]
	\includegraphics[width=\linewidth]{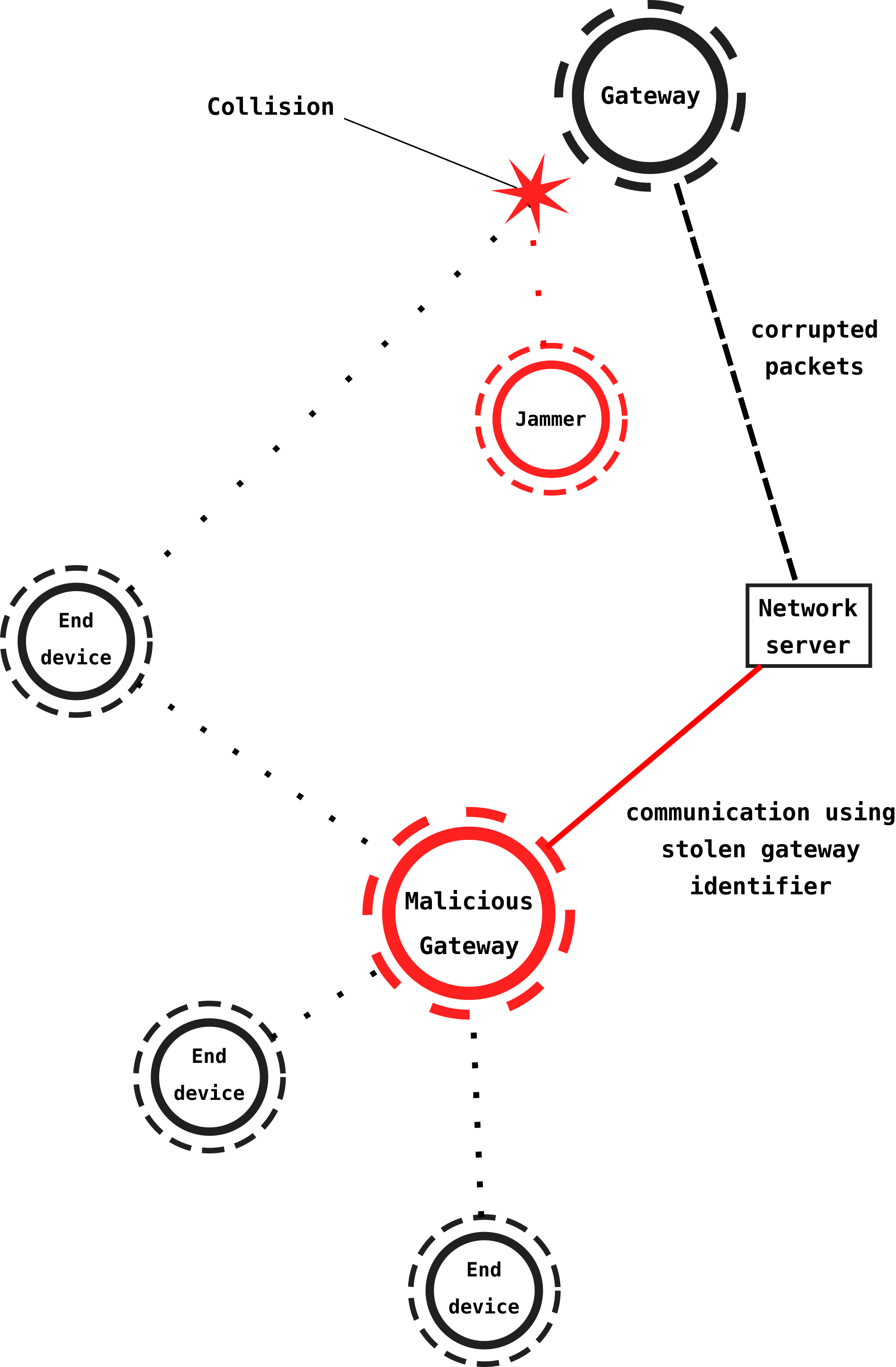}
	\caption{impersonation attack using jamming}
\end{figure}

\section{Countermeasures}

Fortunately a multitude of counteractions can be taken to detect or prevent this kind of impersonation attack. There are a lot of traces that this attack leaves during execution. Also limitations exist, resulting from the need for physical access and certain configurations of gateways and network servers. In the following we describe concepts to detect this attack using intrusion detection systems (IDS). Also five different approaches are presented to either make the attack harder to implement or even completely prevent it.

\subsection{Detection}

With the help of IDS, there are multiple traces to find when impersonation is enacted. First of all, when the gateway is disconnected and the malicious one is switched on, it appears to the server as if the IP address has changed. Depending on the network connection type the gateway uses, this does not necessarily have to be an indication for an impersonation, since Internet service providers tend to change IP addresses of customers in regular intervals\cite{ISPsuperuser}. It can however be a recommending hint to closely observe further behavior. Respectively an advanced network monitoring solution with an endpoint installed on the gateway would immediately show if it is disabled as a change in IP address would not affect the functionality.\\
Changing IP addresses though are a clear sign for an impersonation attack in the scenario of the jammed gateway. As it is assumed that corrupted packets are still sent to the server with "stat" field set to -1, there would be packet loss messages with the original source IP and packets with correct payload coming from the attackers IP. There is no natural scenario explaining this behavior.\\
Investigating communication on the packet forwarder level, there are further inconsistencies in the case of jamming. An increase in frequency of PULL\_DATA messages can be detected, as now two gateways are sending them. If the attacker wants to increase chances for success, this effect is even stronger.\\
Assuming a gateway having no packet loss during normal operation, suddenly the rate of corrupted packets increases to about 50\% if the malicious gateway is turned on simultaneously with the jammer. This in conjunction with other traces can also be a good indication for an attack proceeding.

\subsection{Protection of gateways}

Disabling gateways has to be prevented, therefore physical preparations are necessary. Gateways should be located in secure housings and connected to a power source that is only accessible to trustworthy persons\cite{YangLoRaSec}. If the area around the gateway is not accessible to the public likewise, jamming messages to the gateway will be less successful, as sending a signal with higher power has less impact due to free space attenuation. The network connection has to be stable and immune to disturbances. This suggests that cabled connections should be preferred over wireless, cellular or other radio based options. Also LoRa gateway EUIs should be treated with more care, presenting them on a map or leaving them in open repositories is discouraged.

\subsection{Gateway redundancy}

As soon as one functioning gateway is covering an end device, wormhole and ACK spoofing attacks are rendered ineffective. Messages will be sent and received via the working gateway. LoRaWAN uses sequential counters for plink and downlink messages so there is no problem in the overlapping of gateways as a message that is received again, it is simply dropped. Overlapping gateways in general can help to improve availability and resistance against jamming\cite{SelJam}.

\subsection{Securing connection between gateway and server}

In general it can be recommended to not allow traffic from unregistered gateways. Accepting packets from such is making impersonation unnecessary since a jamming and wormhole attack does suffice in this case.\\
Since with impersonation the same can still be achieved after taking this measure, further steps should be taken to secure the connection. One way is to implement communication secured by IPSec as the attacker in our scenario would have no way of getting hands on the cryptographic keys, it would not make a difference if tunnel or transport mode, authentication header or encapsulated security payload is used. As long as the network server is configured to not accept packets that are not secured with IPSec in some way, the attack is unsuccessful. The same principle applies to virtual private networks on link or transport layer. As long as some kind of functioning authentication is used on these network layers the described impersonation attack fails.

\subsection{Make jamming harder}

As shown by Aras et al. the success of jamming attacks is directly depending on the spreading factor especially in the case of selective jamming\cite{SelJam}. This can be seen in Table 5, which shows the results of the selective jamming and wormhole attack. Using smaller spreading factors leads to shorter LoRa messages which are harder to be recorded or scanned in context of a selective jamming attack. As this is a feature which is used when adapting to environments subject to noise and other disturbances, a balance has to be found in these situations.

\begin{table}[h]
	\begin{tabular}{ l | l} 
		SF & packet loss \\
		\hline
		7 & <0\% \\
		8 & <0\% \\
		9 & 0-95\% \\
		10 & >95\% \\
		11 & >95\% \\
		12 & >95\%\\
	\end{tabular}
	\caption{Capabilities of the Wormhole Jammer setup for
		each SF for a packet size of 37 bytes. Successful jamming: (>95\%), mixed success: (0-95\%), failure to jam: (<0\%) | taken from \cite{SelJam}}
\end{table}

\section{Conclusion}

In this paper we showed that there hypothetically are two possible ways to impersonate a LoRaWAN gateway using the Semtech Packet Forwarder Protocol under the assumption of it being registered to a LoRaWAN in the manner of TTN. These ways involve the disabling of gateways. We do this either via disconnecting the gateway from the power source or its network connection or by jamming messages from end devices on their way to the gateway. We find that the absence of authentication in the protocol and confidentiality of the gateway's unique identifier are making it possible to enact a gateway in a realistic scenario. Traces left during the conduction of the attack are then proposed to be used for detection in intrusion detection systems. We propose effective ways like increasing physical gateway protection, redundancy in gateways and securing the gateway-to-server connection, as well as adjusting transmission parameters to diminish the success of jamming attacks to counter these threats. There are still parameters which we can not examine by interpreting protocol specifications and making use of related works. The questions of how well the attack performs in practice and how network servers choose the receiver of downlink messages remain for experimental evaluation. In case of the jamming approach, the attack also depends on how the network server chooses the logical address of the gateway. This scenario therefore needs tests in real environments as well. There might be similar flaws in other protocols as LoRaWAN is not specifying a protocol for gateway connection so manufacturers are using and developing different protocols. In the interest of interoperability between products of distinct manufacturers it might be helpful if an optional standard is proposed, considering the security risks. Drawbacks resulting from a loss of flexibility would not be present in the optional case. Proposing such a standard however could convince some manufacturers to use it as a default, leaving inexperienced users less vulnerable.

\section*{Acknowledgements}

Parts of this work were created under supervision of Lars Almon from the Secure Mobile Networking Lab (SEEMOO) at Technische Universtit\"at Darmstadt during the Advanced Seminar on Networking, Security, Mobility, and Wireless Communications summer term 2018.

\begin{scriptsize}
\bibliographystyle{abbrv}
\bibliography{bib}  
\end{scriptsize}

\end{document}